\documentstyle[psfig,12pt]{article}
\def\er{{\tilde e}_R}
\def\el{{\tilde e}_L}
\def\nl{{\tilde \nu}_e}
\def\etal{{\it et al.}}
\def\beq{\begin{equation}}
\def\eeq{\end{equation}}
\def\bea{\begin{eqnarray}}
\def\eea{\end{eqnarray}}
\def\ba{\begin{array}}

\def\simgt{\stackrel{>}{{}_\sim}}
\def\ea{\end{array}}
\def\gappeq{\mathrel{\rlap {\raise.5ex\hbox{$>$}}
{\lower.5ex\hbox{$\sim$}}}}
\def\permil{$\%\raise.20ex\hbox{$_0$}}
\def\lappeq{\mathrel{\rlap{\raise.5ex\hbox{$<$}}
{\lower.5ex\hbox{$\sim$}}}}
\begin{document}
\topmargin -1.0cm
\oddsidemargin -0.8cm
\evensidemargin -0.8cm
\pagestyle{empty}
\begin{flushright}
DESY 96-257 \\
CERN-TH/96-352 \\
hep-ph/9612334 
\end{flushright}
\vspace*{5mm}
\begin{center}
{\Large \bf Four-Jet Signal at LEP2 and Supersymmetry}\\
\vspace{1.5cm}
{\large M. Carena$^{a,b}$,
G.F. Giudice$^a$\footnote{On leave of absence from INFN, Sezione di
Padova,
Padua, Italy.}, S. Lola$^a$ and C.E.M. Wagner$^a$}\\
\vspace{0.3cm}
$^a$ Theory Division, CERN\\
Geneva, Switzerland\\
\vspace{0.3cm}
$^b$ Deutsches Elektronen-Synchrotron, DESY\\
Hamburg, Germany\\
\vspace{0.3cm}
\vspace*{2cm}
Abstract
\end{center}
ALEPH has reported a significant excess of four-jet events in the LEP runs 
above the $Z^0$ resonance, which however has not been confirmed by the 
other LEP collaborations. We assume here that this excess corresponds
to a physics signal and try to interpret it in the context of 
supersymmetric models with $R$-parity violation. Associated production
of a left and right selectron can explain all the distinctive features
of the ALEPH data: the value of the cross section, the dijet mass
difference, the absence of bottom quarks in the final state, and the
dijet charge content. Our proposed scenario makes definite predictions,
which can be tested at future LEP runs at higher energies.
\vfill
\begin{flushleft}
CERN-TH/96-352\\
December 1996
\end{flushleft}
\eject
\pagestyle{empty}
\setcounter{page}{1}
\setcounter{footnote}{0}
\pagestyle{plain}

\section{Introduction}

One of the most intriguing and controversial results of the LEP run above
the $Z^0$ resonance has been the excess of four-jet events reported by
ALEPH \cite{jet,rag}. As the other three experimental collaborations working
at LEP do not observe any anomaly in four-jet topologies, 
the resolution of the experimental controversy is a most urgent
issue. All experimental
collaborations are actively working on the question, which hopefully will
be settled by further study and, most importantly, by the new runs at higher
energies. From the theoretical side, we believe that it is important to
investigate if the reported ALEPH data can be interpreted as a consistent
physics signal. At least such a study can be used as a benchmark to compare
the present results with future data at higher $\sqrt{s}$.

The ALEPH four-jet events have been selected from the data 
recorded at centre-of-mass energies between
130 and 172 GeV. An excess is observed in the distribution of the sum of
the two dijet invariant masses contructed by pairing jets with the smallest
dijet mass difference. This distribution shows a peak at $106.1\pm 0.8$
GeV, corresponding to 18 events observed with 3.1 expected from QCD
background \cite{rag}. If interpreted as particle pair production, this
corresponds to a cross section of $2.5\pm 0.7$ pb when only data
with $\sqrt{s}$ in the
range between 130 and 161 GeV are considered, 
and of $1.5\pm 0.8$ pb when all data with $\sqrt{s}$ in the
range between 130 and 172 GeV are considered \cite{rag}. This cross 
section is too large for Higgs 
bosons or for electroweakly-interacting scalar particles, whose productions
are suppressed by a factor $\beta^3$. Here $\beta$ is the final-state
particle velocity in the centre-of-mass, which is,
for the relevant kinematical 
configuration, about 0.6 at
$\sqrt{s}=130$ GeV and 0.8 at $\sqrt{s}=172$ GeV. The inferred value
of the cross section could be accommodated by production
of fermions with electroweak couplings or of scalar particles with 
a substantial colour or multiplicity factor.

The dijet mass difference distribution of the selected 18 events
is consistent with a value 
around 10 GeV \cite{jet,rag}. Combining this with the information
on the dijet mass sum, it can be concluded that the pair-produced particles
should have masses of about 58 and 48 GeV, respectively. Pair-production
of equal-mass particles is disfavoured.

At the moment little information can be extracted 
from angular distributions. From measurements of the ``rapidity-weighted" 
jet charge, a variable that
statistically retains information on the electric charge of the primary
parton \cite{cha}, one concludes \cite{jet}
that the pair-produced particles have a sizeable
charge. Electrically neutral particles are therefore disfavoured. Finally
there is little or no presence of $b$ quarks in the final states 
\cite{jet,rag}. This is
another reason to reject the hypothesis of Higgs-boson production.

In this paper we want to study whether the ALEPH data, assumed here to 
correspond to a 
real physics signal, can be explained by pair production of a left-handed
and a right-handed selectron, each particle
decaying into two quarks, as an effect of $R$-parity violating interactions.
Other interpretations of the four-jet events have already been presented in
the literature \cite{oth}, but to our knowledge this is the first example
of a consistent picture of pair production and decay
of particles with different masses, in agreement with
the results of the ALEPH analysis.

\section{Particle Production}

Let us start by considering the particle-production cross sections. 
Left-handed or right-handed selectrons can be pair-produced at LEP in
any of the channels $\el \el$, $\er \er$, and $\el \er$. The interactions
involved in the production of the pairs $\el \el$ or $\er \er$ always 
require that the incoming electron and positron have opposite helicities
({\it i.e.} collinear spin vectors).
This means that the cross section for scalar particle production has
a $\beta^3$ suppression, corresponding to a 
$p$-wave suppression near threshold.
On the other hand, the
production of $\el \er$ pairs involves the interaction of an electron
and a positron with the same helicity ({\it i.e.} opposite spin vectors),
and the cross section near threshold is proportional to $\beta$, corresponding
to an $s$-wave.

The differential cross section for $\el \er$ production is
\beq
\frac{d\sigma}{dt}(e^+e^-\to\el^+\er^-)=
\frac{d\sigma}{dt}(e^+e^-\to\er^+\el^-)
 =\frac{g^4\tan^4\theta_W}{64\pi s}\sum_{a,b=1}^4 \frac{A_aA_bM_{\chi^0_a}
M_{\chi^0_b}}{(t-M^2_{\chi^0_a})(t-M^2_{\chi^0_b})}~,
\label{cross}
\eeq
\beq
A_a\equiv N_{a1}\left( N_{a1}+\frac{N_{a2}}{\tan\theta_W}\right)~.
\eeq
Here $N_{a1}$ and $N_{a2}$ are the $B$-ino and $W_3$-ino components of
the $a$-th neutralino with mass $M_{\chi^0_a}$. In the limit of a purely
$B$-ino state, only one neutralino contributes to the sum in eq.~(\ref{cross}).
Therefore the $U(1)$ gaugino mass $M_1$ is the most important parameter
entering eq.~(\ref{cross}). For simplicity, we will concentrate
on the case in which the lightest neutralino is a pure $B$-ino. 
Because of the necessary helicity
flip proportional to the gaugino mass, the cross section decreases only
as $M_1^{-2}$, for large $M_1$.

To reproduce the kinematical configuration suggested by the ALEPH data,
we choose $m_{\el} =58$ GeV and $m_{\er} =48$ GeV. The electron sneutrino mass
is then also determined by the weak $SU(2)$ relation
\beq
m_{\nl}^2 = m_{\el}^2+(1-\sin^2\theta_W)\cos 2\beta ~M_Z^2~,
\label{neum}
\eeq
where $\tan\beta$ is the usual ratio of Higgs vacuum expectation values.
We assume here that the sleptons of the second and third
generations are heavier than those of the first one, and cannot have been
produced at LEP. We will comment in sect.~5 on the case in which slepton
masses are universal in flavour.

In fig.~1 we show the cross sections at $\sqrt{s}=130$ and 172 GeV
for the different production
channels $\el \er$, $\el \el$, $\er \er$, $\nl \nl$, and $\chi_1^0
\chi_1^0$ as a function
of the $U(1)$ gaugino mass $M_1$,
in the limit of large $\mu$ and $M_2$.
The cross sections are corrected for
initial-state radiation and have been generated by the numerical code
SUSYXS \cite{man}.
The important result is that, for $M_1$ less than about 100 GeV,
the $\el \er$ production cross section is large, and
consistent with the value suggested by the ALEPH data. Also, 
in the range $M_1=80$--100 GeV, all other
particle--antiparticle production channels 
are quite suppressed. For the
$\el \el$ and $\er \er$ channels, this
is the result of an efficient destructive interference between the
$s$-channel $\gamma$/$Z$ exchange and the $t$-channel neutralino/chargino
exchange. To sufficiently suppress the $\nl$ cross section, we have
to choose $m_{\nl}$ close to the upper bound determined by eq.~(\ref{neum}).
This implies that $\tan \beta$ is close to 1, and the top Yukawa coupling
is quite large. This can be made consistent with the absence of a Landau
pole below the grand unification scale only if some new physics threshold
exists. Indeed, the need of an effective supersymmetry-breaking scale 
$\Lambda_{\rm SUSY}$
much lower than the Planck scale is also suggested, in our scenario,
by the presence of small slepton masses together with larger gaugino masses.
In fact, for very large values of  $\Lambda_{\rm SUSY}$, such hierarchy 
of masses would 
require a large amount of fine-tuning between the 
value of the slepton 
and the gaugino mass parameters at the high energy scale 
(for a recent discussion of the dependence of the 
renormalization group evolution of the scalar mass parameters on 
the effective supersymmetry breaking scale, see ref. \cite{car}). 

The results presented in fig.~1 correspond to the case in which the lightest
neutralino is a pure $B$-ino. Had we assumed unification of gaugino masses,
and values of the higgsino mass $\mu$ not too large, then the cross section
for $\el \er$ production could be larger than what is shown in fig.~1, as a
consequence of the mixing between $B$-ino and $W_3$-ino, see 
eq.~(\ref{cross}). However the
$\nl$ production cross section would also sizeably increase, because of
the constructive interference between chargino and $Z$ exchange contributions.
For instance, for large values of $|\mu|$, and $M_2 = 500 \; (300)$ GeV, the 
charged slepton production cross sections at $\sqrt{s} = 172$ GeV
are not significantly modified, but the sneutrino cross section
is enhanced from 0.4 pb to 0.51 (0.72) pb for $m_{\nl} = 53$ GeV,
while for $m_{\nl} = 58$ GeV the cross section is enhanced from  
0.31 pb to 0.40 (0.57) pb. Hence, values of $M_2 \simgt 500$ GeV
will efficiently suppress the $\nl$ production cross section.

The differential cross section for $\el \er$ production, eq.~(\ref{cross}),
leads to an angular distribution that is different from the usual
scalar-particle pair production via gauge bosons in the $s$-channel
with $d\sigma /dt \propto (ut-m^4)$. In fig.~2
we compare the two distributions as a function of the angle $\theta$ between
the beam direction and one of the two dijet momenta ($0<\theta < \pi /2$).
At the moment, the experimental uncertainties are too large 
for us to distinguish
between the two distributions. If the charge of the primary parton is
identified, one can  measure the forward--backward asymmetry 
${\cal A}_{FB}$ of the dijet
system with a definite charge. In the case of ordinary scalar particle
pair production, the distribution is symmetric in the forward and backward
regions, and ${\cal A}_{FB}=0$. 
This is however not true for the distribution in eq.~(\ref{cross}),
which produces the following integrated forward--backward asymmetry
\beq
{\cal A}_{FB}=\frac{\sqrt{\left[ s-(m_{\el} +m_{\er} )^2\right]
\left[ s-(m_{\el} -m_{\er} )^2\right]}}{s+2 M_1^2-m^2_{\el} -m^2_{\er}} ~.
\label{fbasy}
\eeq
Here we have assumed that the $B$-ino is an approximate mass eigenstate,
and defined the forward and backward regions with respect to initial-
and final-state particles with the same electric charge. For $M_1=80$--100
GeV,
${\cal A}_{FB}$ is large, about 40--60\% at $\sqrt{s}=130$ GeV and
30--50\% at $\sqrt{s}=172$ GeV.

\section{Particle Decay}

In order to generate the four-jet final state from the slepton pair, we 
have to introduce some
$R$-parity violating interaction. The only renormalizable operator that
couples quarks to leptons has the following expression in the superpotential:
\beq
\lambda_{ijk} L_L^i Q_L^j {\bar D}_R^k~.
\label{rp}
\eeq
We have used here a standard notation for lepton and quark chiral superfields,
and denoted the generation indices as $i,j,k$. We assume that one of the
couplings $\lambda_{ijk}$ is much larger than all the others; this
coupling determines the decay mode of the lightest supersymmetric particle.
As we do not want to consider
top or bottom quarks in the final state, we are interested only in the
couplings $\lambda_{1jk}$ with $j,k=1,2$. If in the future
more experimental information
on the flavour content of the jets becomes 
available, we will be able to further 
restrict the choice of the operators.

Our interpretation of the four-jet events as slepton pairs requires that
the $R$-parity violating decay mode has a branching ratio close to 1.
Thus it is important to compare the rate for $\el$ decay into two quarks,
\beq
\Gamma (\el^- \to \bar{u}_j d_k )=\frac{3\lambda_{1jk}^2}{16\pi}~m_{\el}~,
\eeq
with the $R$-parity conserving decay rates. Indeed,
$\el$ can decay into the lightest supersymmetric particle, $\er$, 
through neutralino exchange. In the
approximation $M_{\chi^0}\gg m_{\el} , m_{\er}$, the decay widths are
\beq
\Gamma (\el^- \to e^- e^+ \er^- )=\frac{g^4 \tan^4\theta_W}{3(8\pi)^3}
m^3_{\el} F_1 \left( \frac{m^2_{\er}}{m^2_{\el}}\right)
\sum_{a,b=1}^4 \frac{A_a A_b}{M_{\chi^0_a}M_{\chi^0_b}} ~,
\eeq
\newline
\beq
F_1(x)=(1-x)(1+10x+x^2)+6x(1+x)\ln x~,
\eeq
\newline
\beq
\Gamma (\el^- \to e^- e^- \er^+ )=\frac{g^4 \tan^4\theta_W}{3(16\pi)^3}
m^5_{\el} F_2 \left( \frac{m^2_{\er}}{m^2_{\el}}\right)
\sum_{a,b=1}^4 \frac{A_a A_b}{M^2_{\chi^0_a}M^2_{\chi^0_b}} ~,
\eeq
\newline
\beq
F_2(x)=(1-x)(1-7x-7x^2+x^3)-12x^2\ln x ~.
\eeq
Also, $\el$ can decay into $\nl$ via $W$ exchange
\beq
\Gamma (\el^- \to f \bar{f^\prime} \nl )= N_{ff^\prime} \frac{G_F^2
m_{\el}^5}{3(4 \pi)^3}F_2 \left( \frac{m^2_{\nl}}{m^2_{\el}}\right) ~,
\eeq
Here $N_{ff^\prime}$ is a colour factor, equal to 9, when
summed over the light quarks and 
leptons in the final state. The decay rate for $\el^- \to e^- \nu_e
\bar{\nl}$ can be neglected, as it is suppressed by the chargino mass.

Figure 3 shows the value of $BR(\el^- \to \bar{u}_j d_k )$, as a result
of a phase-space integration in the limit of a purely $B$-ino neutralino,
but with no approximations on the ratio $M_{\chi^0_1}
/m_{\el}$. The $R$-parity violating mode dominates the $\el$ decays for
$\lambda_{1jk}$ larger than few times $10^{-4}$.
These values for the
$R$-parity violating coupling constants are consistent with present bounds,
as we discuss in the following.

Experimental bounds on $\lambda_{1jk}$ depend on the values of the squark
masses, which mediate the effective four-fermion interactions between the
leptons and quarks. We give here the bounds for a typical squark mass
of 300 GeV, although the value of the squark mass does 
not enter into our analysis.
The heavier the squarks are, the weaker the bounds on $\lambda_{1jk}$ become.
{}From charged current universality, one finds $\lambda_{11k}<0.1$ \cite{bar}.
{}From limits on 
$BR(K^+\to \pi^+ \nu \bar \nu )$, one finds $\lambda_{1jk}<0.03$ 
\cite{ber}, although this limit depends on assumptions about the flavour
structure. From radiative contributions to the electron neutrino mass, one
can get significant limits only for $R$-parity violating operators that
involve a third generation index \cite{hal}. From negative searches of
neutrinoless double-$\beta$ decay, one obtains an interesting limit on
$\lambda_{111}<8\times 10^{-3}$ \cite{ger}, and a bound
on the product $\lambda_{121}
\lambda_{112}<3\times 10^{-5}$ \cite{moh}. Experiments at HERA have
set bounds on $\lambda_{1jk}$ of about $10^{-1}$ for squark masses of 
200 GeV; these bounds disappear for values of the squark masses above
300 GeV \cite{her}.
The only problematic constraint
comes from cosmological considerations about the survival of a baryon
asymmetry created at the very early stages of the Universe, which requires
$\lambda_{1jk}<10^{-7}$ \cite {cos}. However, this limit does not apply 
to cosmological models with low-temperature baryogenesis, and can also
be evaded under certain conditions \cite{cli}.

We therefore conclude that there is a large range of $\lambda_{1jk}$ values,
consistent with present bounds on $R$-parity violation,
in which $\el$ decays almost entirely into a quark pair. This is true,
although $\el$ is not the lightest supersymmetric particle, because the
$R$-parity violating 
two-body decay is more important than kinematically suppressed three-body
decay modes.

In our scenario, $\er$, the lightest supersymmetric particle, 
does not participate in the $R$-parity violating interaction in 
eq.~(\ref{rp}), which involves only quarks and left-handed leptons.
Therefore the $\er$ decay will occur either 
via the small mixing $\phi$ between $\er$ and $\el$,
\beq
\Gamma (\er^- \to \bar{u}_j d_k )=\frac{3}{16\pi} 
\lambda_{1jk}^2 \sin^2\phi ~ m_{\er} ~,
\eeq
or via virtual neutralino
and $\el$ exchange,
\beq
\Gamma (\er^- \to e^+e^- \bar{u}_j d_k ) =\frac{3g^4 \tan^4\theta_W
\lambda_{1jk}^2 m_{\er}^3}{4(4\pi)^5} ~G_1\left( \frac{m^2_{\el}}{m^2_{\er}}
\right) \sum_{a,b=1}^4 \frac{A_aA_b}{M_{\chi^0_a}M_{\chi^0_b}} ~,
\label{3bod1}
\eeq
\newline
\beq
G_1(x)=\frac{(x-1)}{6}(4x^2+25x+1)\ln \left( \frac{x}{x-1}\right)
-\frac{(12x^2+123x+13)}{18}+x(3x+2){\rm Li}\left( \frac{1}{x}\right) ~,
\eeq
\newline
\beq
\Gamma (\er^- \to e^-e^- u_j {\bar d}_k ) =\frac{3g^4 \tan^4\theta_W
\lambda_{1jk}^2 m_{\er}^5}{(8\pi)^5}~ G_2\left( \frac{m^2_{\el}}{m^2_{\er}}
\right) \sum_{a,b=1}^4 \frac{A_aA_b}{M_{\chi^0_a}^2M_{\chi^0_b}^2} ~,
\label{3bod2}
\eeq
\newline
\beq
G_2(x)=\frac{(x-1)}{6}(5x^3-27x^2-15x+1)\ln \left( \frac{x}{x-1}\right)
-\frac{(60x^3-354x^2-460x+43)}{72}-6x^2{\rm Li}\left( 
\frac{1}{x}\right) .
\eeq

In fig.~4 we show the $BR$($\er \to \bar{u}_j d_k $) as a function of the
mixing angle $\phi$, in the limit of a purely $B$-ino neutralino. 
Again, although eqs.~(\ref{3bod1})--(\ref{3bod2}) have
been derived in the approximation $M_{\chi^0_a} \gg m_{\el} ,m_{\er}$,
the results plotted in fig.~4 follow from a numerical integration
of phase space with no restrictive assumptions.

The mixing angle $\phi$ is related to the higgsino mass $\mu$ and to
the trilinear coupling $A$ by the relation
\beq
\sin \phi \simeq \frac{m_e(A-\mu \tan\beta )}{m^2_{\el} -m^2_{\er}}=
\left( \frac{A-\mu \tan \beta}{200 ~{\rm GeV}}\right) ~10^{-4}~.
\eeq
Therefore the most plausible values for $\sin \phi$ lie in the range
around $10^{-4}$. From fig.~4 we then infer that, in this range, 
$\er$ dominantly decays into two jets. In conclusion, although $\er$
does not participate in the $R$-parity violating interaction, the small
left--right mixing ensures that the preferred $\er$ decay mode is into
a quark pair, rather than into a phase-space suppressed four-body
final state.

The $R$-parity violating coupling $\lambda_{1jk}$ does not influence
the $\er$ decay branching ratio, as long as it is non-vanishing.
It determines however the $\er$ lifetime, which is
\beq
\tau_{\er} = \left( \frac{10^{-4}}{\sin \phi}\right)^2
\left( \frac{10^{-2}}{\lambda_{1jk}}\right)^2~2\times 10^{-13}~{\rm s}~.
\eeq
For the relevant kinematical configuration, this correspond to a 
decay vertex displacement of about
\beq
d_{\er} =
\sqrt{\frac{[ s-(m_{\el} +m_{\er} )^2]
[ s-(m_{\el} -m_{\er} )^2 ]}{4sm^2_{\er}}} ~\tau_{\er}\simeq
\left( \frac{10^{-4}}{\sin \phi}\right)^2
\left( \frac{10^{-2}}{\lambda_{1jk}}\right)^2 ~50 \;
{\rm to} \; 80~\mu{\rm m}~.
\eeq
Depending on the values of $\sin \phi$ and $\lambda_{1jk}$, this could be
measured at LEP.

\section{Prospects for LEP Searches at Higher Energies}

The best testing ground for the plausibility of the ALEPH data will
come with the new LEP runs at higher $\sqrt{s}$. If the ALEPH signal
is real and our interpretation correct, we should expect pair production
of $\el \el$, $\er \er$, $\el \er$, and $\nl \nl$ 
with the rates shown in table 1. The preferred sneutrino decay mode
is $\nl \to {\bar d}_j d_k$, as the $R$-parity conserving decay modes
$\nl \to \nu_e e^\pm \er^\mp$ are suppressed by phase space and by the
small mixing between $B$-ino and $W_3$-ino. Therefore
all slepton production 
processes correspond to four-jet events, although the peaks
in the distributions of the sum and difference of dijet masses depend
on the process. 

Charginos are expected to be too heavy to be produced at LEP, even
if gaugino
mass unification holds. There is however a chance to observe the 
lightest neutralino $\chi^0_1$, if the parameter $M_1$ is in the lower 
part 
of the allowed range. The $B$-ino state $\chi^0_1$ decays
into two jets and an electron with more than 80\% probability, or else into
two jets and a neutrino. The relevant
production cross sections are also shown in table 1.

Finally there are very good prospects for the discovery
of the Higgs boson.
Within the supersymmetric model with minimal Higgs structure, the low
values of $\tan \beta$ assumed here imply that the lightest Higgs boson
has Standard Model-like couplings and a mass, coming almost entirely
from radiative corrections, roughly below 80 GeV \cite{rep}.

\section{The Case of Flavour Universality}

As we have mentioned before, in our analysis we have assumed that sleptons
of the second and third generations are heavier than those of the first.
This assumption is not inconsistent with the strong bounds on individual
lepton number conservation, derived from $\mu \to e \gamma$ and similar
processes. An approximate lepton flavour conservation can be the result
of an alignment between leptons and sleptons, as a consequence of 
additional global symmetries \cite{nir} or of a dynamical principle \cite{noi}.

Let us suppose now that the slepton supersymmetry-breaking masses are
universal in flavour.
Because of the mixing effect, we find that the lightest smuon and stau 
are lighter than $\er$ by an amount 
\beq
\Delta_{{\tilde \mu}{\tilde e}}\simeq \frac{(A-\mu \tan \beta )^2 m_\mu^2}{2
m_{\er} (m^2_{\el} -m^2_{\er} )} = \left( \frac{A-\mu \tan 
\beta }{200~{\rm GeV}}
\right)^2 ~4~{\rm MeV}~,
\eeq
\beq
\Delta_{{\tilde \tau}{\tilde e}}\simeq \frac{(A-\mu \tan \beta )^2 m_\tau^2}{2
m_{\er} (m^2_{\el} -m^2_{\er} )} = \left( \frac{A-\mu \tan 
\beta }{200~{\rm GeV}}
\right)^2 ~1~{\rm GeV}~.
\eeq
Thus the mainly right-handed stau is the lightest supersymmetric particle.

Because of the absence of $t$-channel contributions, pair productions of
smuons, staus, and their corresponding sneutrinos do not suffer from
destructive interference and have relatively large cross sections.
Some indicative numbers are the following:
For $m_{\tilde{\mu}_R} = 48$ GeV,
the  $\tilde{\mu}_R \tilde{\mu}_R$ production cross section
is $0.55$ pb for $\sqrt{s}=130$ GeV and
$0.54$ pb for $\sqrt{s}=172$ GeV.
For $m_{\tilde{\tau}_R} = 47$ GeV,
the  $\tilde{\tau}_R \tilde{\tau}_R$ production cross section
is $0.59$ pb for $\sqrt{s}=130$ GeV and
$0.57$ pb for $\sqrt{s}=172$ GeV.
For $m_{\tilde{\mu}_L} = 58$ GeV,
the  $\tilde{\mu}_L \tilde{\mu}_L$ production cross section
is $0.18$ pb for $\sqrt{s}=130$ GeV and
$0.43$ pb for $\sqrt{s}=172$ GeV.
For $m_{\tilde{\tau}_L} = 59$ GeV,
the  $\tilde{\tau}_L \tilde{\tau}_L$ production cross section
is $0.14$ pb for $\sqrt{s}=130$ GeV and
$0.41$ pb for $\sqrt{s}=172$ GeV.
Finally, for $m_{\tilde{\nu}} = 58 \; (53) $ GeV,
the  $\tilde{\nu} \tilde{\nu}$ production cross section
is $0.21 \; (0.49)$ pb for $\sqrt{s}=130$ GeV and
$0.31 \; (0.4) $ pb for $\sqrt{s}=172$ GeV.

The simultaneous presence of $R$-parity violating interactions with
$\lambda_{ijk}\ne 0$, for different values of the index $i$, is 
severely constrained by lepton flavour-transition processes like
$\mu \to e \gamma$. We are then led to assume that the
second- and third-generation sleptons do not participate in the $R$-parity
violating interaction, and consequently their decays have to involve 
real or virtual $\er$, $\el$, or $\nl$. Their signatures are therefore
two jets accompanied by soft leptons or small amounts of missing energy.
The presence of the leptons and/or neutrinos is a necessary feature of
the transition between different generations of sleptons.
Since such events have not been reported by any of the LEP experimental
collaborations, we believe that the case of universality has to be rejected.

\section{Conclusions}

In this paper we have assumed that the controversial ALEPH excess
of four-jet events corresponds
indeed to a physics signal and we have interpreted it in the context of a
supersymmetric model with $R$-parity violation.
If we consider non-universal mass terms for gauginos and for sleptons
with different flavours, we find that $\el \er$ production can reproduce
the four-jet events, while the production of
other associated supersymmetric particles
occurs at a much lower rate. Within an acceptable range of
$R$-parity violating
couplings, both $\el$ and $\er$ naturally
have a decay branching ratio into two quarks very close to 1.

Our model is compatible with all discernible features emerging
from the ALEPH data: the value of the cross section, the dijet mass
difference, the absence of bottom quarks in the final state, and the jet
charge content. It also predicts a specific angular distribution and a
large forward--backward asymmetry in the jet charge. LEP runs at higher
energies will be able to confirm or rule out this scenario.

We wish to thank P.~Janot for very useful discussions. 
We also acknowledge conversations with J.~Marcos and P.~Morawitz.
The work of S.L. is funded by a Marie Curie 
Fellowship (TMR-ERBFMBICT-950565).

\def\ijmp#1#2#3{{\it Int. Jour. Mod. Phys. }{\bf #1~}(19#2)~#3}
\def\pl#1#2#3{{\it Phys. Lett. }{\bf B#1~}(19#2)~#3}
\def\zp#1#2#3{{\it Z. Phys. }{\bf C#1~}(19#2)~#3}
\def\prl#1#2#3{{\it Phys. Rev. Lett. }{\bf #1~}(19#2)~#3}
\def\rmp#1#2#3{{\it Rev. Mod. Phys. }{\bf #1~}(19#2)~#3}
\def\prep#1#2#3{{\it Phys. Rep. }{\bf #1~}(19#2)~#3}
\def\pr#1#2#3{{\it Phys. Rev. }{\bf D#1~}(19#2)~#3}
\def\np#1#2#3{{\it Nucl. Phys. }{\bf B#1~}(19#2)~#3}
\def\mpl#1#2#3{{\it Mod. Phys. Lett. }{\bf #1~}(19#2)~#3}
\def\arnps#1#2#3{{\it Annu. Rev. Nucl. Part. Sci. }{\bf
#1~}(19#2)~#3}
\def\sjnp#1#2#3{{\it Sov. J. Nucl. Phys. }{\bf #1~}(19#2)~#3}
\def\jetp#1#2#3{{\it JETP Lett. }{\bf #1~}(19#2)~#3}
\def\app#1#2#3{{\it Acta Phys. Polon. }{\bf #1~}(19#2)~#3}
\def\rnc#1#2#3{{\it Riv. Nuovo Cim. }{\bf #1~}(19#2)~#3}
\def\ap#1#2#3{{\it Ann. Phys. }{\bf #1~}(19#2)~#3}
\def\ptp#1#2#3{{\it Prog. Theor. Phys. }{\bf #1~}(19#2)~#3}

\pagebreak

\begin{table}[p]
\begin{center}
\caption{Predictions for new particle production at future LEP runs.
We have taken $m_{\el}=58$ GeV and $m_{\er}=48$ GeV.}
\vspace*{1 cm}
\begin{tabular}{|c|c|c|c|}
\hline
Process & Particle mass & $\sigma$ at $\sqrt{s}=186$ GeV 
& $\sigma$ at $\sqrt{s}=195$ GeV \\
 & [GeV] & [pb] & [pb] \\
\hline
\hline
 & & & \\
$\el \er$ & $M_1=80$           & 1.46 & 1.37 \\
$\el \er$ & $M_1=100$          & 1.33 & 1.27\\
$\el \el$ & $M_1=80$           & 0.20 & 0.19\\
$\el \el$ & $M_1=100$          & 0.23 & 0.22\\
$\er \er$ & $M_1=80$           & 0.42 & 0.47\\
$\er \er$ & $M_1=100$          & 0.20 & 0.22\\
$\nl \nl$ & $m_{\nl}=53$       & 0.36 & 0.33\\
$\nl \nl$ & $m_{\nl}=58$       & 0.29 & 0.28\\
$\chi^0_1 \chi^0_1$ & $M_1=80$ & 0.62 & 0.80\\
 & & & \\
\hline
\end{tabular}
\end{center}
\end{table}

\pagebreak
\begin{figure}
\centerline{
\psfig{figure=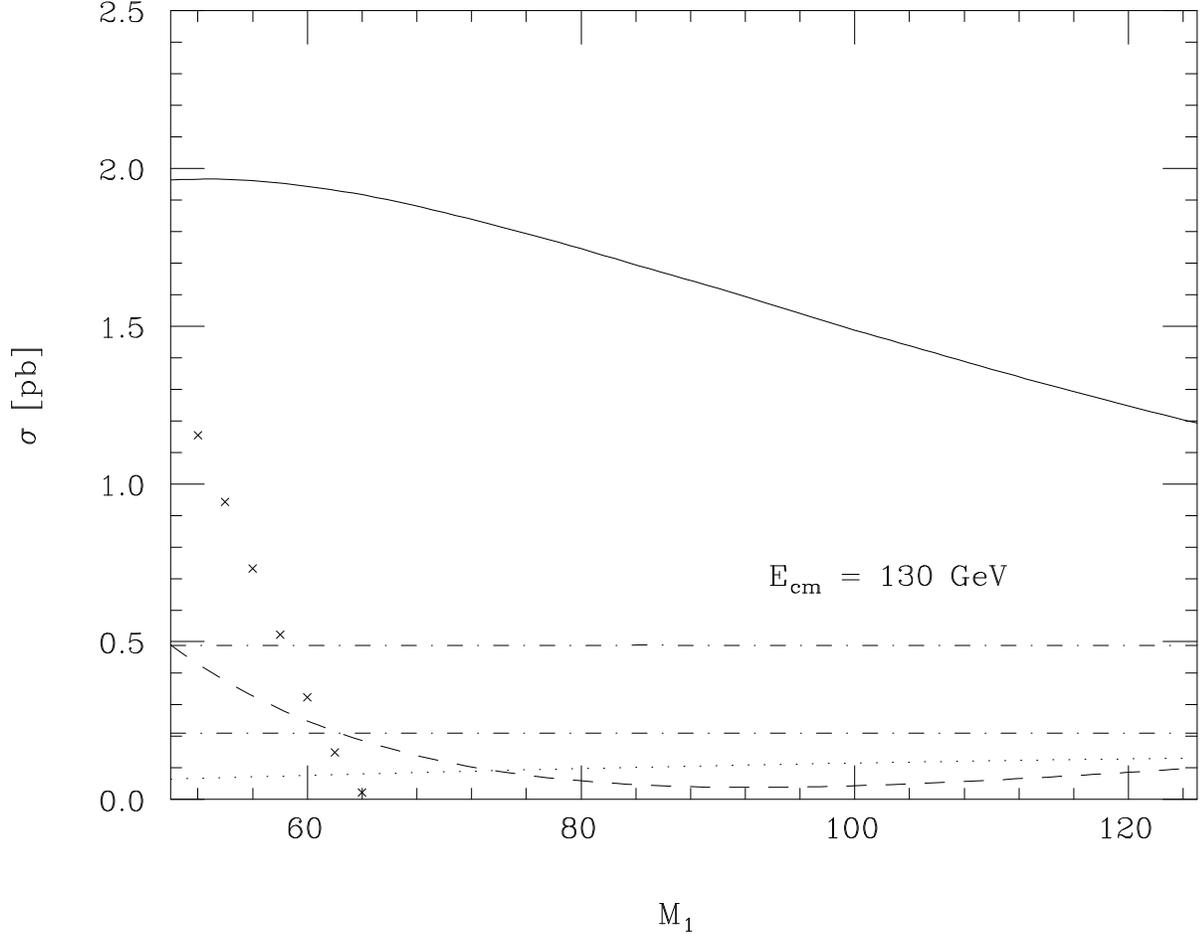,width=20cm,height=15cm,angle=90}}
\caption[0]{a)
Cross sections at $\sqrt{s}=130$ 
for the production
channels: $\el \er$ (solid lines), $\el \el$ (dotted lines), $\er \er$
(dashed lines), $\nl \nl$ (dot-dashed lines), and $\chi_1^0
\chi_1^0$ (crosses) as a function
of the $U(1)$ gaugino mass $M_1$. 
We have taken $m_{\el}=58$ GeV and $m_{\er}=48$ GeV. The upper dot-dashed
line corresponds to $m_{\nl}=53$ GeV and the lower one to $m_{\nl}=58$ GeV.}
\end{figure} 

\begin{figure}
\setcounter{figure}{0}
\centerline{
\psfig{figure=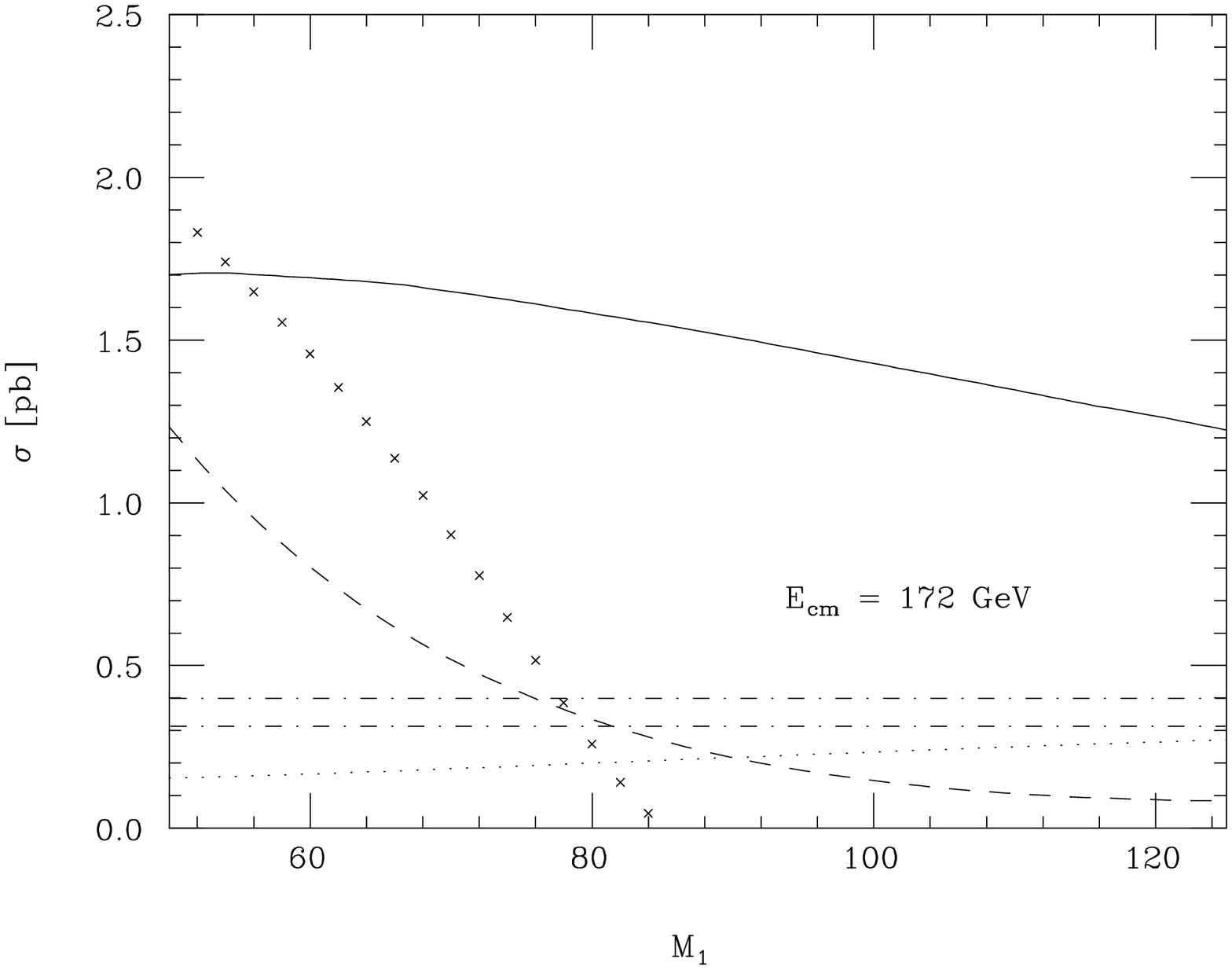,width=20cm,height=15cm,angle=90}}
\caption[0]{b) The same as Fig.1.a but at $\sqrt{s} = 172$ GeV.}
\end{figure}

\begin{figure}
\centerline{
\psfig{figure=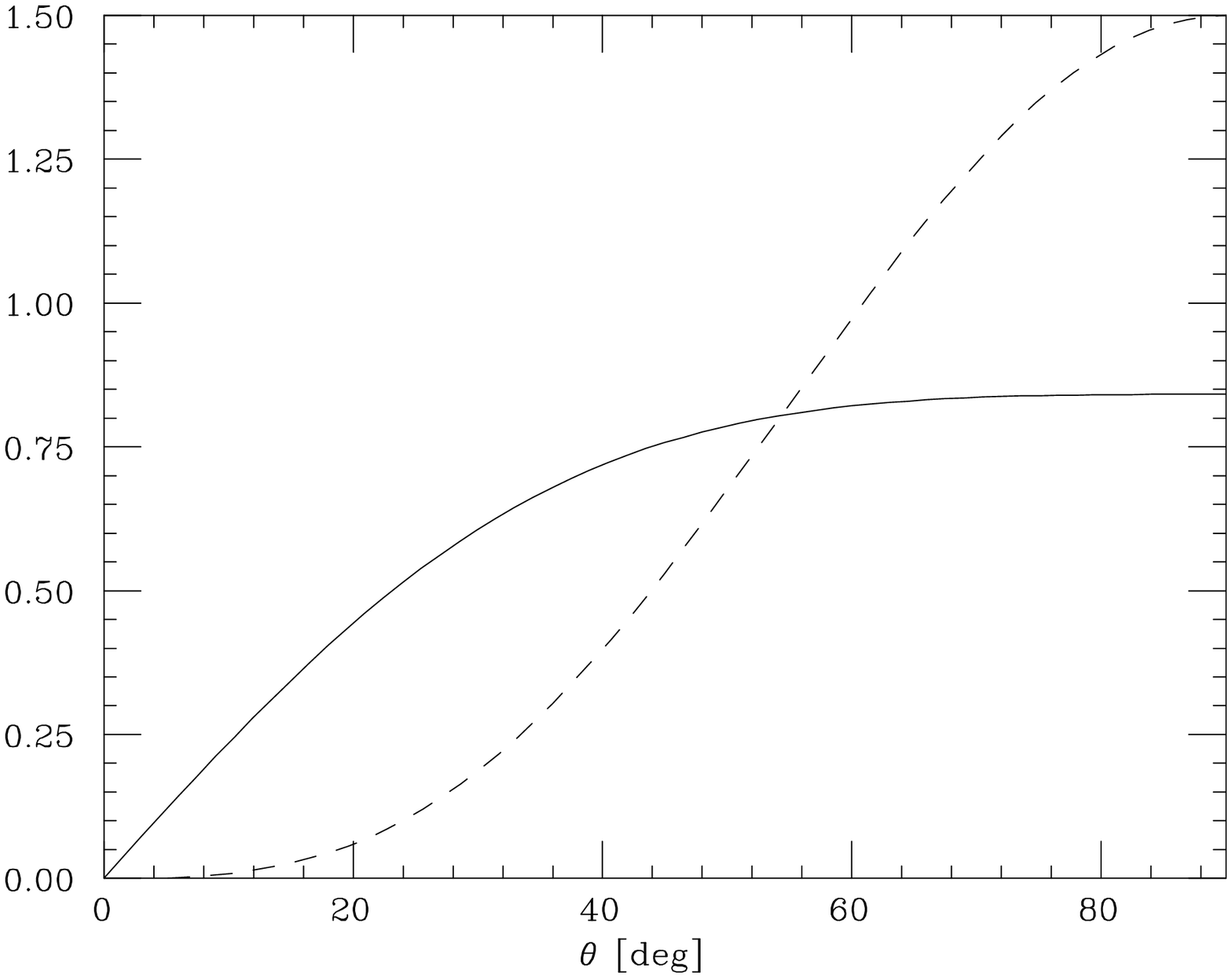,width=20cm,height=15cm,angle=90}}
\caption[0]{Distributions in the angle $\theta$ between the
dijet momentum and the beam direction for  scalar-particle pair
production via gauge bosons in the $s$-channel (dashed line) 
and for $\el \er$ production (solid line).
The curves are normalized so that their integrals over $0<\theta <\pi /2$
are equal to 1.
We have taken $m_{\el}=58$ GeV, $m_{\er}=48$ GeV, $M_1=90$ GeV, and
$\sqrt{s}=136$ GeV.}
\end{figure}

\begin{figure}
\centerline{
\psfig{figure=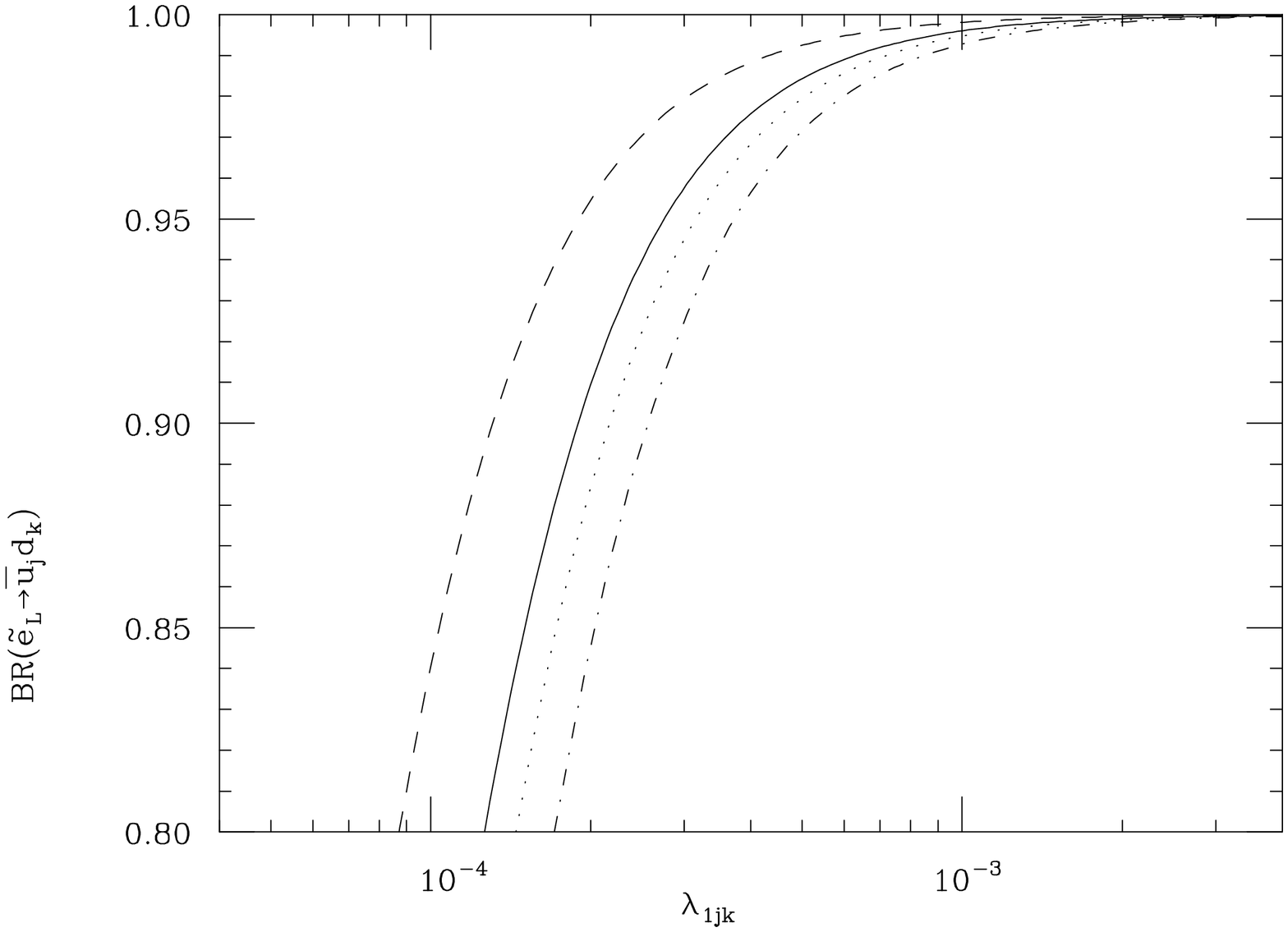,width=20cm,height=15cm,angle=90}}
\caption[0]{Branching ratio for $\el \to \bar{u}_j d_k $, as
a function of the $R$-parity violating coupling $\lambda_{1jk}$.
We have taken $m_{\el}=58$ GeV, $m_{\er}=48$ GeV, 
and $M_1=80$ GeV, $m_{\nl}=53$ (dot-dashed line);
$M_1=80$ GeV, $m_{\nl}=58$ (dotted line);
$M_1=100$ GeV, $m_{\nl}=53$ (solid line);
$M_1=100$ GeV, $m_{\nl}=58$ (dashed line).}
\end{figure}

\begin{figure}
\centerline{
\psfig{figure=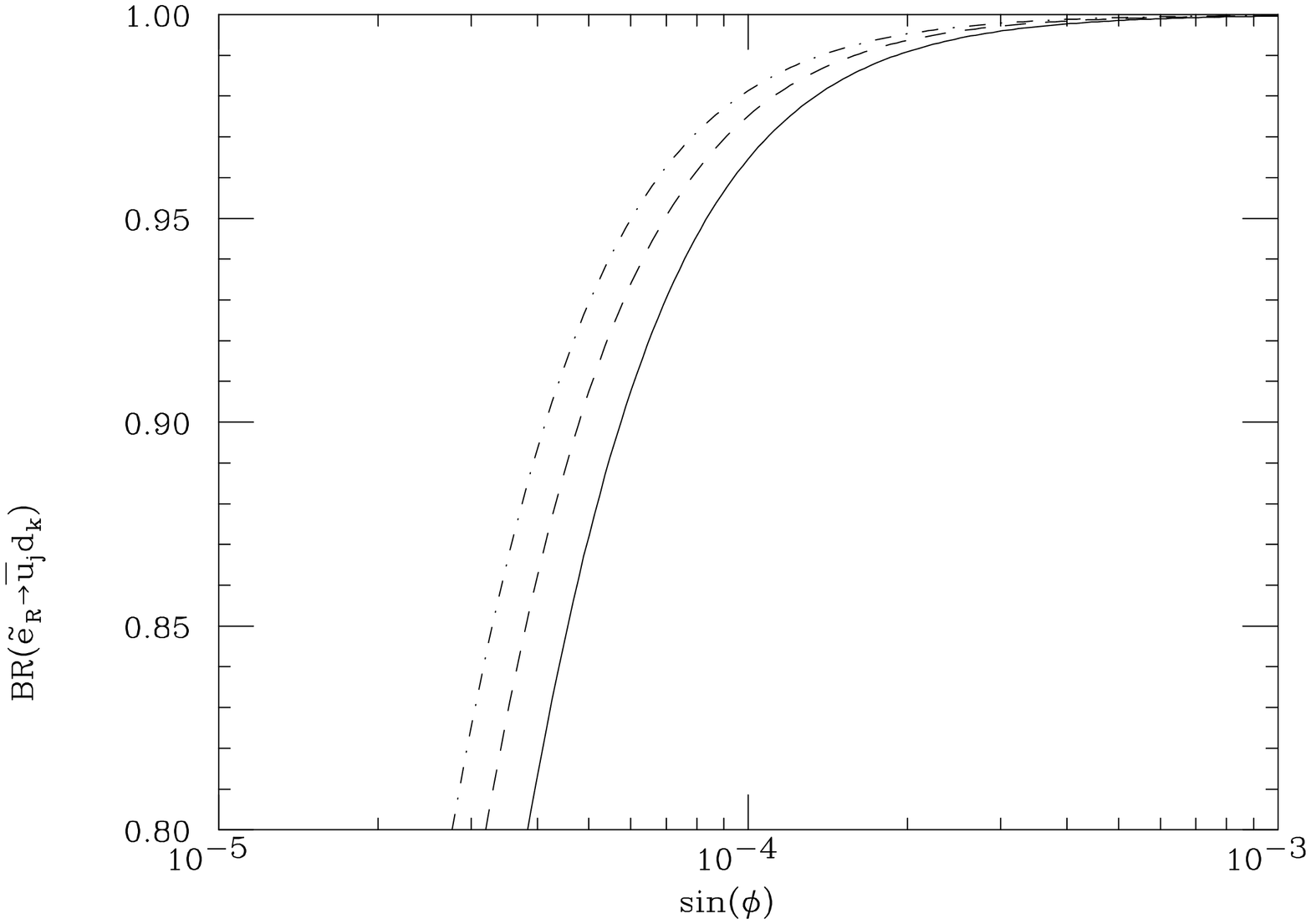,width=20cm,height=15cm,angle=90}}
\caption[0]{Branching ratio for $\er \to \bar{u}_j d_k $, as
a function of the left--right mixing angle $\phi$.
We have taken $m_{\el}=58$ GeV, $m_{\er}=48$ GeV, $M_1=80$ GeV (solid
line), 90 GeV (dashed line), or 100 GeV (dot-dashed line).}
\end{figure}

\end{document}